\documentclass[11pt]{article}
\usepackage{amsmath,epsfig,sint}
\usepackage[american]{babel}
\font\sixrm=cmr6

\def\rmd{{\rm d}}

\def\rme{{\rm e}}
\def\rmO{{\rm O}}

\def\bfx{{\bf x}}

\def\defeq{\mathrel{\mathop=^{\rm def}}}
\def\proof{\noindent{\sl Proof:}\kern0.6em}

\def\frac#1#2{\hbox{$#1\over#2$}}
\def\dual{\mathstrut^*\kern-0.1em}

\def\lvec#1{\setbox0=\hbox{$#1$}
    \setbox1=\hbox{$\scriptstyle\leftarrow$}
    #1\kern-\wd0\smash{
    \raise\ht0\hbox{$\raise1pt\hbox{$\scriptstyle\leftarrow$}$}}
    \kern-\wd1\kern\wd0}
\def\rvec#1{\setbox0=\hbox{$#1$}
    \setbox1=\hbox{$\scriptstyle\rightarrow$}
    #1\kern-\wd0\smash{
    \raise\ht0\hbox{$\raise1pt\hbox{$\scriptstyle\rightarrow$}$}}
    \kern-\wd1\kern\wd0}

\def\nab#1{{\nabla_{#1}}}
\def\nabstar#1{\nabla\kern-0.5pt\smash{\raise 4.5pt\hbox{$\ast$}}
               \kern-4.5pt_{#1}}

\def\drvstar#1{\partial\kern-0.5pt\smash{\raise 4.5pt\hbox{$\ast$}}
               \kern-5.0pt_{#1}}

\def\momp#1#2{
    \setbox0=\hbox{${#1}$}\setbox1=\hbox{${#1}_{#2}$}
    {#1}_{#2}\kern-\wd1\kern\wd0
    \smash{\raise4.5pt\hbox{$\scriptscriptstyle +$}}}
\def\momm#1#2{
    \setbox0=\hbox{${#1}$}\setbox1=\hbox{${#1}_{#2}$}
    {#1}_{#2}\kern-\wd1\kern\wd0
    \smash{\raise4.5pt\hbox{$\scriptscriptstyle -$}}}
\def\mompm#1#2{
    \setbox0=\hbox{${#1}$}\setbox1=\hbox{${#1}_{#2}$}
    {#1}_{#2}\kern-\wd1\kern\wd0
    \smash{\raise4.5pt\hbox{$\scriptscriptstyle \pm$}}}
\def\smomp#1#2{
    \setbox0=\hbox{${#1}$}\setbox1=\hbox{${#1}_{#2}$}
    {#1}_{#2}\kern-\wd1\kern\wd0
    \smash{\raise3pt\hbox{$\scriptscriptstyle +$}}}
\def\smomm#1#2{
    \setbox0=\hbox{${#1}$}\setbox1=\hbox{${#1}_{#2}$}
    {#1}_{#2}\kern-\wd1\kern\wd0
    \smash{\raise3pt\hbox{$\scriptscriptstyle -$}}}
\def\smompm#1#2{
    \setbox0=\hbox{${#1}$}\setbox1=\hbox{${#1}_{#2}$}
    {#1}_{#2}\kern-\wd1\kern\wd0
    \smash{\raise3pt\hbox{$\scriptscriptstyle \pm$}}}
\def\si{\kern1pt{\rm si}}
\def\co{\kern1pt{\rm co}}

\def\Nf{N_{\rm f}}
\def\psibar{\bar{\psi}}
\def\psiL{\psi_{\rm L}}
\def\psiR{\psi_{\rm R}}
\def\psibarL{\psibar_{\rm L}}
\def\psibarR{\psibar_{\rm R}}

\def\psiprime{\psi\kern1pt'}
\def\psibarprime{\psibar\kern1pt'}
\def\rhoprime{\rho\kern1pt'}
\def\rhobar{\bar{\rho}}
\def\rhobarprime{\rhobar\kern1pt'}
\def\rhobartilde{\kern2pt\tilde{\kern-2pt\rhobar}}
\def\rhobartildeprime{\kern2pt\tilde{\kern-2pt\rhobar}\kern1pt'}

\def\zetabar{\bar{\zeta}}
\def\zetaprime{\zeta\kern1pt'}
\def\zetabarprime{\zetabar\kern1pt'}
\def\zetar{\zeta_{\raise-1pt\hbox{\sixrm R}}}
\def\zetabarr{\zetabar_{\raise-1pt\hbox{\sixrm R}}}

\def\phiimpr{\phi_{\kern0.5pt\hbox{\sixrm I}}}

\def\dirac#1{\gamma_{#1}}
\def\diracstar#1#2{
    \setbox0=\hbox{$\gamma$}\setbox1=\hbox{$\gamma_{#1}$}
    \gamma_{#1}\kern-\wd1\kern\wd0
    \smash{\raise4.5pt\hbox{$\scriptstyle#2$}}}

\def\f1{f_1}

\def\h1{h_1}

\def\SUtwo{{\rm SU(2)}}

\def\tr{\,\hbox{tr}\,}

\def\opprime#1{\setbox0=\hbox{${\cal O}$}\setbox1=\hbox{${\cal O}_{\rm #1}$}
    {\cal O}_{\rm #1}\kern-\wd1\kern\wd0
    \smash{\raise4.5pt\hbox{\kern1pt$\scriptstyle\prime$}}\kern1pt}

\def\ophatprime#1{\setbox0=\hbox{$\widehat{\cal O}$}
    \setbox1=\hbox{$\widehat{\cal O}_{\rm #1}$}
    \widehat{\cal O}_{\rm #1}\kern-\wd1\kern\wd0
    \smash{\raise4.5pt\hbox{\kern1pt$\scriptstyle\prime$}}\kern1pt}

\def\bopprime#1{\setbox0=\hbox{${\cal O}$}\setbox1=\hbox{${\cal O}_{\rm #1}$}
    {\cal L}_{\rm #1}\kern-\wd1\kern\wd0
    \smash{\raise4.5pt\hbox{\kern1pt$\scriptstyle\prime$}}\kern1pt}

\def\blagprime#1{\setbox0=\hbox{${\cal B}$}\setbox1=\hbox{${\cal B}_{#1}$}
    {\cal B}_{#1}\kern-\wd1\kern\wd0
    \smash{\raise5.2pt\hbox{\kern1pt$\scriptstyle\prime$}}\kern1pt}

\def\gr{g_{{\hbox{\sixrm R}}}}

\def\muq{\mu_{\rm q}}
\def\mq{m_{\rm q}}

\def\mr{m_{{\hbox{\sixrm R}}}}
\def\mur{\mu_{{\hbox{\sixrm R}}}}
\def\mc{m_{\rm c}}

\def\za{Z_{\rm A}}
\def\zv{Z_{\rm V}}
\def\zp{Z_{\rm P}}

\def\Za{\za}
\def\Zv{\zv}
\def\Zp{\zp}

\def\msbar{{\rm \overline{MS\kern-0.05em}\kern0.05em}}

\newcommand{\bes}{\begin{eqnarray}}
\newcommand{\ees}{\end{eqnarray}}

\begin{document}
\begin{titlepage}
\begin{flushright}
   CERN-TH/2000-384 \\
   MPI-PhT/2000-51\\
   Bicocca-FT-0027\\
   NYU-TH/00/09/13\\
   December 2000
\end{flushright}
\vskip 0.5 cm
\begin{center}
  {\Large\bf Lattice QCD with a chirally twisted mass term\\[1.5ex]
   }
\end{center}
\vskip 0.5 cm
\begin{figure}[h]
\begin{center}
\epsfig{figure=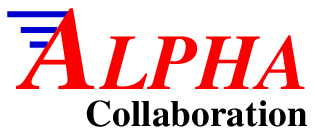} 
\end{center}
\end{figure}
\begin{center}
{\large Roberto Frezzotti$^{\scriptscriptstyle a}$,
        Pietro Antonio Grassi$^{\scriptscriptstyle b}$,\\[1ex]
        Stefan Sint$^{\scriptscriptstyle c}$
    and Peter Weisz$^{\scriptscriptstyle d}$}
\end{center}
\vskip 2.3ex
\begin{flushleft}
$^{\scriptstyle a}$ Universit\`a di Milano-Bicocca, 
Dipartimento di Fisica, Piazza della Scienza~3,
I--20126 Milano, Italy\\[1ex]
$^{\scriptstyle b}$ Department of Physics, New York University,
4 Washington Place, New York, NY 10003, U.S.A.\\[1ex]
$^{\scriptstyle c}$ CERN, Theory Division,
CH--1211 Geneva 23, Switzerland\\[1ex]
$^{\scriptstyle d}$ Max-Planck-Institut f\"ur Physik,
F\"ohringer Ring 6, D--80805 M\"unchen, Germany
\end{flushleft}
\begin{center} 
 {\bf Abstract}
\end{center}
\vskip 0.7ex
Lattice QCD with Wilson quarks and a chirally twisted mass term
represents a promising alternative regularization of QCD,
which does not suffer from unphysical fermion zero modes.
We show how the correlation functions of the renormalized theory are
related to the theory with a standard parameterization of the mass term.
In particular we discuss the conditions under which these
relations take the same form as obtained from naive continuum considerations.
We discuss in detail some applications and comment on potential
benefits and problems of this framework.

\vfill
\eject

\end{titlepage}

\section{Introduction}

Lattice QCD with Wilson quarks~\cite{Wilson} is widely used 
to compute hadronic observables and matrix elements from first
principles. It is a gauge invariant regularization with
an ultra-local action and an exact global flavour symmetry, 
but all axial symmetries are explicitly broken by the Wilson term. 
The latter fact is usually not considered a fundamental problem, 
as chiral symmetry can be restored by introducing appropriate
counterterms. Well-known examples are the additive 
quark mass renormalization, the renormalization 
of the non-singlet axial current and the mixing pattern of the 
$\Delta S=2$ effective weak Hamiltonian. The corresponding 
scale independent renormalization constants 
can be determined both in perturbation theory 
and non-perturbatively, by imposing continuum chiral 
Ward identities as normalization conditions~\cite{Bochicchio_et_al,paperIV}.

As a result chiral symmetry is restored up
to cutoff effects, and the problem has thus been 
solved from a field theoretical point of view.
However, the absence of an exact chiral symmetry
has further consequences in practical applications, 
as it implies that the Wilson-Dirac operator 
is not protected against zero modes.
This is not a problem in principle, as the functional integral
over Grassmann variables cannot diverge. After integration 
over the quark fields, a small eigenvalue of the Wilson-Dirac 
operator appears both in the fermionic determinant and 
in the quark propagators entering the
correlation functions. Fermi statistics then implies that the
limit of a vanishing eigenvalue is always regular. 
Despite this fact, numerical simulations with 
some of the standard algorithms may still experience technical problems.
In particular, one may suspect that accidental zero modes
are at the origin of long autocorrelation times 
which have been observed in numerical simulations 
with the hybrid Monte Carlo algorithm~\cite{HMCtrouble}. 

A conceptual problem arises in the so-called quenched 
approximation, which consists in neglecting the fermionic determinant. 
The contribution of a small eigenvalue to a fermionic 
correlator is then not balanced by the determinant, leading to 
large fluctuations in some of the observables which completely compromise
the ensemble average~\cite{paperIII}. 
Gauge field configurations where this happens are called
``exceptional'' and various recipes of how to deal with them
have appeared in the literature~\cite{BardeenI-Schierholz}. 
Strictly speaking, the quenched approximation with Wilson fermions
is ill-defined, as the absence of zero modes is only guaranteed
for rather heavy quarks. We emphasise that this problem is
common to all lattice regularizations with Wilson type fermions.
However, its practical relevance in a given physical
situation depends on all the details of the chosen lattice action,
and on the statistics one would like to achieve in numerical simulations. 
For example, with non-perturbatively O($a$) improved Wilson 
quarks~\cite{SW,paperIII} and the standard Wilson plaquette action, 
the problem is felt when the quark masses become 
somewhat lighter than the strange quark's mass~\cite{ALPHAexceptional}. 
If the quark mass is further decreased, 
the frequency of (near-) exceptional configurations 
strongly increases. As the problem becomes even more pronounced
with increasing lattice volume~\cite{paperIII}, 
it is clear that the approach to the chiral limit 
with  Wilson type quarks is limited by the zero mode problem 
rather than by finite volume effects. 

To solve the aforementioned practical and conceptual problems 
we propose to add a non-standard mass term to the  
(improved) Wilson quark action. The lattice Dirac operator
for two quark flavours then reads
\begin{equation}
   D_{\rm tmQCD}=D_{W}+m_0+i\muq\gamma_5\tau^3,
\label{D_tmQCD}
\end{equation}
where $D_W$ denotes the massless Wilson-Dirac operator,
$m_0$ is the standard bare quark mass, and $\muq$ is referred
to as twisted mass parameter. It couples to a term with a
non-trivial flavour structure ($\tau^3$ is a Pauli matrix 
acting in flavour space), and protects the Dirac 
operator against zero modes, independently of the background gauge field.
This lattice action has previously  appeared in a different 
context~\cite{Aoki}, and it has already been proposed as a regulator 
for exceptional configurations, implying, however, 
a limiting procedure $\muq \rightarrow 0$ at the end~\cite{BardeenII}. 
In contrast, our proposal, first presented in~\cite{lat99},
is to interpret this theory as an alternative regularization
of QCD with two mass degenerate quarks. We will refer to 
it as QCD with a chirally twisted mass term, or 
twisted mass QCD (tmQCD) for short. Indeed,  
in the classical continuum limit an axial rotation 
of the quark and anti-quark fields relates 
the tmQCD action to the standard QCD action. Furthermore 
the axial rotation of the fundamental fields induces 
a mapping between composite fields. One may hence think 
of this transformation as a change of variables 
which leaves the physical content of the theory unchanged.

In this paper we want to demonstrate in which sense 
these classical considerations can be elevated to 
a relation between the renormalized correlation functions of tmQCD
and standard QCD. To this end we first regularize both
tmQCD and standard QCD using Ginsparg-Wilson 
quarks (sect.~2). In this framework the
bare correlation functions can be related by a change of variables 
in the functional integral. Renormalization
will be discussed in sect.~3, in particular we identify 
renormalization schemes which preserve the relations between 
the bare correlation functions.
Based on universality, it will then be clear how to proceed 
if the regularization does not respect chiral symmetry, and we 
discuss in detail the case of Wilson quarks~(sect.~4).
We conclude with a few remarks concerning current and 
future work on tmQCD~(sect.~5).

\section{Twisted mass QCD and Ginsparg Wilson quarks}

We start with classical continuum considerations, and then discuss
the regularization with Ginsparg-Wilson fermions. In particular
we use a formulation which hides the lattice peculiarities
as much as possible, so that naive continuum relations
carry over essentially unchanged to the lattice regularized theory. 

\subsection{Classical continuum theory}

We consider the continuum limit 
of the twisted mass QCD 
action\footnote{We adhere to the convention that $\gamma$-matrices
are hermitian, $\{\gamma_\mu,\gamma_\nu\}=2\delta_{\mu\nu}$, 
with $\mu,\nu,\ldots$ ranging from $0$ to $3$, and we set 
$\gamma_5=\gamma_0\gamma_1\gamma_2\gamma_3$.},
\begin{equation}
  S_{\rm F}[\psi,\psibar] =\int \rmd^4x\,\psibar\left(D\kern-7pt\slash
                +m+i\muq\gamma_5\tau^3\right)\psi, 
\label{tmQCDcont}
\end{equation}
where $D_\mu=\partial_\mu+G_\mu$ 
denotes the covariant derivative in a given gauge field $G_\mu$,
and $\tau^3$ is the third Pauli matrix acting in flavour space.
The axial transformation
\begin{equation}
 \psi'    =\exp(i \alpha\gamma_5\tau^3/2)\psi,\qquad
 \psibar' =\psibar\exp(i \alpha\gamma_5\tau^3/2),
 \label{axial}
\end{equation}
leaves the form of the action invariant, and merely 
transforms the mass parameters
\begin{eqnarray}
     m'&=&m\cos(\alpha)+\muq\sin(\alpha), \label{mprime}\\
  \muq'&=&-m\sin(\alpha)+\muq\cos(\alpha).\label{muqprime}
\end{eqnarray}
In particular, the standard action with $\muq'=0$ is obtained 
if the rotation angle $\alpha$ satisfies the relation
\begin{equation}
 \tan{\alpha}=\muq/m.
 \label{angle}
\end{equation}
Chiral symmetry of the massless action leads to the definition
of the Noether currents,
\begin{equation}
  A_\mu^a = \psibar\gamma_\mu\gamma_5{{\tau^a}\over{2}}\psi,\qquad
  V_\mu^a = \psibar\gamma_\mu{{\tau^a}\over{2}}\psi,
  \label{currents_local}
\end{equation}
which are only partially conserved at non-zero mass parameters.
More precisely, the so-called PCAC and 
PCVC relations take the form
\begin{eqnarray}
  \partial_\mu A_\mu^a &=& 2m P^a+i\muq \delta^{3a}S^0, \label{PCAC}\\
  \partial_\mu V_\mu^a &=& -2\muq\,\varepsilon^{3ab} P^b,\label{PCVC}
\end{eqnarray}
where we have defined the pseudo-scalar and scalar densities,
\begin{equation}
  P^a  = \psibar\gamma_5{{\tau^a}\over{2}}\psi,\qquad
  S^0  = \psibar\psi.
\label{densities}
\end{equation}
The axial transformation of the quark and anti-quark
fields induces a transformation of the 
composite fields. For example, the rotated axial and vector currents read,
\begin{xalignat}{2}
  {A'}_\mu^a &\equiv \psibar'\gamma_\mu\gamma_5{{\tau^a}\over{2}}\psi'
  \!\!\!\!\!\!\!\!&=\;
  \begin{cases}
    \cos(\alpha)A_\mu^a + \varepsilon^{3ab}\sin(\alpha) V_\mu^b
     & \text{$(a=1,2)$},\\
    A_\mu^3 & \text{$(a=3)$},
  \end{cases} \label{axial_current_rot}\\
  {V'}_\mu^a &\equiv \psibar'\gamma_\mu{{\tau^a}\over{2}}\psi' &=\;
  \begin{cases}
    \cos(\alpha)V_\mu^a + \varepsilon^{3ab}\sin(\alpha) A_\mu^b
     & \text{$(a=1,2)$},\\
    V_\mu^3 & \text{$(a=3)$},
  \end{cases}    
\end{xalignat}
and similarly, the rotated pseudo-scalar and scalar densities are
given by
\begin{eqnarray}
  {P'}^a &=&
  \begin{cases}
     P^a
     & \text{$(a=1,2)$},\label{axial_density_rot}\\
    \cos(\alpha) P^3+i\sin(\alpha)\frac12 S^0 & \text{$(a=3)$},
  \end{cases} \\
   {S'}^0 &=& \cos(\alpha) S^0 +2i\sin(\alpha)P^3.
   \label{scalar_density_rot}
\end{eqnarray}
It is easy to verify that the rotated currents and densities 
satisfy the PCAC and PCVC relations~(\ref{PCAC},\ref{PCVC}),
with the transformed parameters $m'$~(\ref{mprime}) 
and $\muq'$~(\ref{muqprime}). 
In particular these relations assume their standard form,
\begin{equation}
  \partial_\mu {A'}^{a}_\mu = 2m' {P'}^{a},\qquad
  \partial_\mu {V'}^{a}_\mu = 0,
\end{equation}
if $\alpha$ is related to the mass parameters as in eq.~(\ref{angle}).

Finally, we note that the tmQCD and standard QCD 
actions are exactly related by the transformation~(\ref{axial})
and therefore share all the symmetries. However,
in the chirally twisted basis the symmetry transformations 
may take a somewhat unusual form. 
For example, a parity transformation is realized by
\begin{eqnarray}
   \psi(x)    &\longrightarrow& 
               \gamma_0\exp(i\alpha\gamma_5\tau^3)\psi(x_0,-\bfx),\\
   \psibar(x) &\longrightarrow& 
               \psibar(x_0,-\bfx)\exp(i\alpha\gamma_5\tau^3)\gamma_0,
\end{eqnarray}
and similar expressions can be obtained for the isospin and 
the remaining discrete symmetries. It is then straightforward
to infer the behaviour of composite fields under these transformations.


\subsection{Ginsparg-Wilson quarks}

We now replace continuous Euclidean space time by a 
hyper-cubic lattice with spacing $a$, and we choose 
some standard lattice action for the gauge fields. The precise
choice will not be important in the following, but for definiteness
we may take Wilson's original plaquette action.
As for the quark fields we assume that the lattice Dirac
operator satisfies the Ginsparg-Wilson relation~\cite{GW},
\begin{equation}
  D \gamma_5+\gamma_5 D = a D \gamma_5 D.
 \label{GWr}
\end{equation}
This relation arises naturally in the construction of 
fixed point actions~\cite{Hasenfratz}, and 
an explicit solution for $D$ has been given by Neuberger~\cite{Neuberger}.
In the following we assume that
$D$ is a local operator~\cite{locality} which satisfies eq.~(\ref{GWr})
and has the conjugation property $D^\dagger=\gamma_5 D\gamma_5$.
It then follows that the matrix~\cite{LuscherII}, 
\begin{equation}
  \hat\gamma_5\defeq \gamma_5(1-aD),
\end{equation}
is hermitian and unitary. The massless action of a quark doublet
\begin{equation}
   S_{\rm F}= a^4\sum_{x} \psibar D \psi,
\end{equation}
has a global $\SUtwo\times\SUtwo$ invariance~\cite{LuscherI} 
which can be parameterised
by the transformation 
\begin{eqnarray}
    \psi' & = & \exp(i\omega_{\sixrm V}^a\tau^a/2) 
                \exp(i\omega_{\sixrm A}^a\hat\gamma_5\tau^a/2)\psi, \nonumber\\
    \psibar' & = &\psibar\,\,\exp(i\omega_{\sixrm A}^a\gamma_5\tau^a/2) 
                \exp(-i\omega_{\sixrm V}^a\tau^a/2).
\label{chiral_transf}
\end{eqnarray}
Here $\omega_{\sixrm V,A}^a$ ($a=1,2,3$) are real parameters and
$\tau^a$ are the Pauli matrices acting on the flavour indices. 
Note that these transformations have the same form as in the
continuum, except for the appearance of $\hat\gamma_5$. To mask
this difference we follow refs.~\cite{Niedermayer_98,LuscherII} 
and define left handed fields
\begin{equation}
 \psiL = \frac12(1-\hat\gamma_5)\psi, \qquad
 \psibarL = \psibar\frac12(1+\gamma_5),
\end{equation}
and analogously the right handed fields with the complementary 
projectors. The massless action splits into left and
right handed parts due to the identity
\begin{equation} 
  \psibar D \psi = \psibarL D \psiL + \psibarR D \psiR,
\end{equation}
and the  transformation rules for the chiral fields
are exactly as in the continuum. In particular it is 
straightforward to construct composite fields which 
transform among each other under a chiral transformation. 
Examples are the isosinglet scalar and the isovector pseudo-scalar
densities which may be defined through
\begin{eqnarray}
  S^0 &\equiv& \psibarL\psiR+\psibarR\psiL = \psibar (1-\frac12 aD)\psi
 \label{Szero}\\[1ex]
  P^a &\equiv& \psibarL{{\tau^a}\over{2}}\psiR-\psibarR{{\tau^a}\over{2}}\psiL
  =\psibar \gamma_5{{\tau^a}\over{2}} (1-\frac12 aD)\psi.
 \label{Pa} 
\end{eqnarray}
One finds
\begin{equation}
 {S'}^0\equiv S^0[\psi',\psibar'] = 
 \cos(\omega_A)S^0  + 2 i\sin(\omega_A)u_A^a P^a,
\end{equation}
where $\omega_A$ denotes the modulus of the vector 
$(\omega_A^1,\omega_A^2,\omega_A^3)$, and $u_A^a=\omega^a_A/\omega_A$ is a 
unit vector. Similarly, ${P'}^a$ can be expressed as a linear
combination of $S^0$ and $P^a$, and we 
have thus found a multiplet of bare lattice operators
transforming under the lattice symmetry in the same way
as their continuum counterparts. 
To define the Ginsparg-Wilson regularization of tmQCD
we now use the scalar and pseudo-scalar densities for the mass terms,
i.e.~we write
\begin{equation}
  S_{\rm F} = a^4\sum_x\left[\psibar D\psi +m S^0 + 2i\muq P^3\right].
 \label{tmQCDaction}
\end{equation}
From the preceding discussion it is then clear that
the transformation
\begin{equation}
  \psi'=\exp(i\alpha\hat\gamma_5\tau^3/2)\psi,\qquad
  \psibar'=\psibar\exp(i\alpha\gamma_5\tau^3/2),
 \label{varchange}
\end{equation}
does not change the form of the action, as it corresponds to 
the special choices $\omega_V^a=0$ and $\omega_A^a=\alpha\,\delta^{3a}$
in eq.~(\ref{chiral_transf}). Its effect can thus 
be absorbed in a change of the parameters
as in eqs.~(\ref{mprime},\ref{muqprime}), and the standard 
QCD action is recovered if $\alpha$ satisfies the relation~(\ref{angle}).

\subsection{Functional integral}

We are now prepared to define the functional integral 
of lattice regularized tmQCD with Ginsparg-Wilson
quarks. Denoting the sum of the Wilson plaquette and the fermionic
action by $S$, the partition function is defined by
\begin{equation}
  {\cal Z}=\int D[\psibar,\psi] D[U]\,\rme^{\displaystyle -S}.
\end{equation}
The physical content of the theory can be extracted 
from its correlation functions, i.e.~$n$-point functions of the form
\begin{equation}
  \langle O(x_1,\ldots,x_n)\rangle = {\cal Z}^{-1}
  \int D[\psi,\psibar] D[U]\,\rme^{\displaystyle -S} 
   O(x_1,\ldots,x_n),
\end{equation}
where $O(x_1,\ldots,x_n)$ is a product of 
local gauge invariant composite fields which are localised
at the space time points $x_1,\ldots,x_n$. In the following we
will sometimes indicate the functional dependence upon the 
quark and anti-quark fields in square brackets, i.e.~we 
set $O\equiv O[\psi,\psibar]$.

If one restricts the theory to a finite physical 
space time volume, e.g.~by imposing periodic boundary conditions,
the functional integral at fixed lattice spacing
is a well defined finite dimensional integral, 
and it is then possible to perform a change of variables
of the form~(\ref{varchange}). As the transformation matrices are 
special unitary, the Jacobian is unity. Due to the form invariance
of the action the whole effect is a 
transformation of the parameters in the action 
and of the composite fields, viz.
\begin{equation}
   \left\langle O(x_1,\dots,x_n)\right\rangle_{(m,\muq)} 
 = \left\langle O'(x_1,\dots,x_n)\right\rangle_{(m',\muq')}.
   \label{change_of_variables}
\end{equation}
Here the primed field is implicitly defined through
\begin{equation}
   O'[\psi',\psibar'] = O[\psi,\psibar],
\end{equation}
with the transformed quark and anti-quark fields in~eq.~(\ref{varchange}).
The subscript of the expectation values
reminds us of the transformations of the parameters in
the action according to eqs.~(\ref{mprime},\ref{muqprime}). 
Eq.~(\ref{change_of_variables}) is an exact identity and
defines the starting point for the statement of 
the aforementioned equivalence between the renormalized theories.

In order to prepare the ground for the discussion of 
the renormalization procedure we proceed a little further.
First we notice that the combination
\begin{equation}
   (m')^2 + (\muq')^2= m^2+\muq^2,
\end{equation}
is left invariant by the chiral rotation, so 
that it is convenient to use polar mass coordinates,
\begin{equation}
   m    = M\cos(\alpha),\qquad
   \muq = M\sin(\alpha),
 \label{polar}
\end{equation}
with radial mass $M$ and the angle $\alpha$ chosen 
according to eq.~(\ref{angle}). Second, as the bare theory has 
an exact chiral symmetry, it is useful to decompose 
operators in multiplets which transform
irreducibly under the chiral flavour symmetry transformation.
We will denote the irreducible components of the 
composite fields by $\phi^{(r)}_{kA}(x)$,
where $r$ labels the representation, $k$ identifies the multiplet
and $A$ labels the members of the multiplet. The  general 
$\SUtwo\times\SUtwo$ transformation (\ref{chiral_transf}) is then represented
by a matrix $R^{(r)}$, viz.
\begin{equation}
 \phi^{(r)}_{kA}[\psi',\psibar'] 
 = R^{(r)}_{AB}(\mbox{\boldmath $\omega_{\rm V}$};
                \mbox{\boldmath $\omega_{\rm A}$})
                \phi^{(r)}_{kB}[\psi,\psibar] .
\end{equation}
Simple examples are the multiplets of 
the scalar and pseudo-scalar densities, $(\frac12 S^0,iP^a)$, 
and of the currents 
$(\psibarL\gamma_\mu \frac{\tau^a}{2}\psiL,
  \psibarR\gamma_\mu \frac{\tau^a}{2}\psiR)$.

Without loss of generality we may restrict attention to  
$n$-point functions of such fields, 
and the identity~(\ref{change_of_variables}) then takes the form
\begin{eqnarray}
  \lefteqn{\left\langle 
    \phi^{(r_1)}_{k_1A_1}(x_1)\cdots\phi^{(r_n)}_{k_nA_n}(x_n)
  \right\rangle_{(M,\alpha)}   = }\nonumber\\
  &&\left\{\prod_{i=1}^n R_{A_iB_i}^{(r_i)}(-\alpha')\right\}
  \left\langle
    \phi^{(r_1)}_{k_1B_1}(x_1)\cdots\phi^{(r_n)}_{k_nB_n}(x_n)
  \right\rangle_{(M,\alpha-\alpha')},  
 \label{identity_bare}
\end{eqnarray}
where we have used polar mass coordinates in the subscript,
and the shorthand notation
\begin{equation}
 R^{(r)}(\alpha)\equiv R^{(r)}\left(\mbox{\boldmath $0$}; 0,0,\alpha\right),
\end{equation}
for the rotation induced by the axial U(1) transformation~(\ref{varchange}).
Note that we may keep the angle $\alpha'$ independent of $\alpha$,
and with the choice $\alpha'=\alpha$ the r.h.s.~of 
eq.~(\ref{identity_bare}) consists of correlation 
functions with the standard QCD parameterisation of the mass term.

\subsection{Ward identities}

Symmetries in Quantum Field Theories are usually expressed
in terms of the Ward identities which follow
from infinitesimal symmetry transformations. 
This has the advantage  that identities are obtained between correlation
functions which are defined within the same theory, 
even in the presence of explicit breaking terms.
To derive the Ward identities we start by introducing
the space-time dependent variations of the quark
and anti-quark fields,
\begin{xalignat}{2}
   \delta^a_{\rm A}(\omega)\psi(x) & = 
   \omega(x)\frac{\tau^a}{2}(\hat\gamma_5\psi)(x),& 
      \delta^a_{\rm A}(\omega)\psibar(x) & = 
   \psibar(x)\gamma_5\frac{\tau^a}{2}\omega(x), \\[1ex]       
    \delta^a_{\rm V}(\omega)\psi(x) &= 
   \omega(x)\frac{\tau^a}{2}\psi(x),&  
      \delta^a_{\rm V}(\omega)\psibar(x) &= 
   -\psibar(x)\frac{\tau^a}{2}\omega(x),       
   \label{local_variation} 
\end{xalignat}
with some real-valued function $\omega(x)$. The action
on arbitrary composite fields is defined by treating
the variations like ordinary derivatives. The variation of the
action~(\ref{tmQCDaction}) then yields
\begin{eqnarray}
  \delta^a_{\rm A}(\omega)S &=& - a^4 \sum_{x}\omega(x)
   \left\{
   \partial_\mu^\ast A_\mu^a(x)- 2m P^a(x)-i\muq \delta^{3a}S^0(x)
   \right\},\\
  \delta^a_{\rm V}(\omega)S &=& - a^4 \sum_{x}\omega(x)
   \left\{
    \partial_\mu^\ast V_\mu^a(x) + 2\muq \varepsilon^{3ab} P^b(x)
   \right\},
\end{eqnarray}
with the pseudoscalar and scalar densities as defined above,
and where the divergences of the symmetry currents are given by
\begin{eqnarray}
  \partial_\mu^\ast A_\mu^a(x)&=&\left(1-\frac{am}{2}\right)
   \left\{\psibar(x)\frac{\tau^a}{2} (D\hat\gamma_5\psi)(x)
         -(\psibar D)(x)\frac{\tau^a}{2}(\hat\gamma_5\psi)(x)
   \right\}
  \label{axialdiv}\\[1ex]
  && \mbox{}+\frac{i}{4} a\muq\left\{
    \psibar(x)\tau^3\tau^a (D\psi)(x)
   -(\psibar D\hat\gamma_5)(x)\tau^3\tau^a
    (\hat\gamma_5\psi)(x)\right\}, \nonumber\\[1ex]
  \partial_\mu^\ast V_\mu^a(x)&=&\left(1-\frac{am}{2}\right)
   \left\{\psibar(x)\frac{\tau^a}{2} (D \psi)(x)
         -(\psibar D)(x)\frac{\tau^a}{2}\psi(x)
   \right\}
  \label{vectordiv}\\[1ex]
  && \mbox{}+\frac{i}{4} a\muq\left\{
    \psibar(x)\tau^3\tau^a (D\hat\gamma_5\psi)(x)
   -(\psibar D\hat\gamma_5)(x)\tau^3\tau^a
    \psi(x)\right\}. \nonumber
\end{eqnarray}
Note that these expressions vanish exactly 
upon summation over $x$, so that the existence of the currents is guaranteed
by the lattice Poincar\'e lemma~\cite{LuscherIII}.
The symmetry currents themselves will not be
needed in the following. Explicit expressions for the massless case 
can be found in~\cite{Kikukawa}, for example, 
but one should keep in mind that the integration of 
eqs.~(\ref{axialdiv},\ref{vectordiv}) is ambiguous by terms 
with vanishing divergence. 

The Ward identities now take the generic form
\begin{equation}
   \left\langle \left(\delta^a_{\rm X}(\omega) S\right) 
      {\cal O}\right\rangle = 
   \left\langle \delta^a_{\rm X}(\omega) {\cal O} \right\rangle, 
   \quad {\rm X}= {\rm A,V},  
  \label{Wardomega}
\end{equation}
where ${\cal O}$ denotes some product of local composite fields.
This is an exact identity, as the space-time dependent
change of variables can be made in the functional integral. 
However, the variation of the composite fields does not exactly 
transform members of a given multiplet among each
other\footnote{We thank L.~Giusti
for drawing our attention to this problem. Although this
was known in the literature~(cf.~e.g.~\cite{Kikukawa}) it has 
been overlooked by us in an earlier version of this paper.}. Rather there
are extra terms, which arise due to the non-trivial space-time
structure of $\hat\gamma_5$.
For instance, the axial variation of the pseudo-scalar density 
is given by
\begin{equation}
  \delta^a_{\rm A}(\omega)P^b(y) = \frac12 \delta^{ab}\omega(y)S^0(y)
   + \frac{1}{8}a\,\psibar(y)\tau^b\tau^a\gamma_5
  \left([\omega,D]\hat\gamma_5\psi\right)(y).
  \label{varP}
\end{equation}
We now first assume that $\omega$ is non-zero only in a single lattice
point $x\neq y$.
Eq.~(\ref{varP}) then reduces to
\begin{equation}
   \delta^a_{\rm A}(\omega)P^b(y) =  -\frac{1}{8}a\,\omega(x)\psibar(y)
  \tau^b\tau^a\gamma_5 D(y,x)\left(\hat\gamma_5\psi\right)(x),
  \label{x_neq_y}
\end{equation}
where $D(y,x)$ denotes the kernel of the Ginsparg-Wilson Dirac operator.
Locality of the Ginsparg-Wilson action then implies that the r.h.s.~of 
eq.~(\ref{x_neq_y}) is exponentially small as long as the
distance between $x$ and $y$ is large in lattice
units. We find that this structure is generic and therefore conclude
that the bare PCAC and PCVC relations hold up to exponentially
small corrections, provided all operators
in the correlation function keep a large distance 
(in lattice units) from the space-time region where the action is varied.

We now assume that $\omega$ is constant in a space-time region $R$,
\begin{equation}
  \omega(x)= \left\{ 
 \begin{split} 1, \quad & \text{if $x\in R$}\\  0, \quad &\text{otherwise,} 
  \end{split}
  \right.
\end{equation}
which leads to 
\begin{equation}
   \delta^a_{\rm A}(\omega)P^b(y) = \left\{
   \begin{split}
    \frac{1}{2} \delta^{ab}S^0(y)+\frac{1}{8}a^5\sum_{x\notin R} 
    \psibar(y)\gamma_5 \tau^b\tau^aD(y,x)(\hat{\gamma_5}\psi)(x),
    \quad \text{if $y \in R$},\\
     -\frac{1}{8}a^5\sum_{x\in R} 
    \psibar(y)\gamma_5 \tau^b\tau^aD(y,x)(\hat{\gamma_5}\psi)(x),
    \quad \text{if $y\notin R$},\\
    \end{split}
                                   \right.
 \label{extra}
\end{equation}
Again, locality of the action 
implies that the extra terms are exponentially small in both cases, 
provided one has 
\begin{eqnarray}
 \min_{x\in R}||x-y||    &\gg & a,\quad  \text{if $y \notin R$}
 \label{cond1}\\
 \min_{x\notin R}||x-y|| &\gg & a,\quad  \text{if $y \in R   $}.
 \label{cond2}
\end{eqnarray}
Furthermore, we note that the above example of the pseudoscalar
density is generic as the difference to naive continuum
considerations always consists of terms 
involving the lattice Dirac operator.
Up to exponentially small corrections, 
the bare Ward identities~(\ref{Wardomega}) can
therefore be written in the continuum like form
\begin{eqnarray}
 \lefteqn{a^4\sum_{x\in R}\left\langle\left( \partial_\mu^\ast A_\mu^a(x) 
  - 2m P^a(x)-i\muq \delta^{3a}S^0(x)\right) 
  \prod_{i=1}^n \phi_{k_iA_i}^{(r_i)}(x_i)\right\rangle=}
  \label{AWI}\\
 &&-i\sum_{i\vert x_i\in R} \left(T_A^{(r_i)}\right)^a_{A_iB}\left\langle 
    \phi_{k_1A_1}^{(r_1)}(x_1)\cdots\phi_{k_iB}^{(r_i)}(x_i)\cdots
    \phi_{k_nA_n}^{(r_n)}(x_n)\right\rangle, \nonumber\\[2ex]
 \lefteqn{a^4\sum_{x\in R}\left\langle\left( \partial_\mu^\ast V_\mu^a(x) 
  + 2\muq \varepsilon^{3ab} P^b(x)\right) 
  \prod_{i=1}^n \phi_{k_iA_i}^{(r_i)}(x_i)\right\rangle=}
   \label{VWI}\\
&& -i\sum_{i\vert x_i\in R}\left(T_V^{(r_i)}\right)^a_{A_iB}\left\langle 
    \phi_{k_1A_1}^{(r_1)}(x_1)\cdots\phi_{k_iB}^{(r_i)}(x_i)\cdots
    \phi_{k_nA_n}^{(r_n)}(x_n)\right\rangle,\nonumber
\end{eqnarray}
where we have assumed the conditions~(\ref{cond1},\ref{cond2}) to
hold with $y$ replaced by $x_i$, for all $i=1,\ldots,n$.
We have omitted the subscript $(M,\alpha)$ of the correlation functions,
and the expansion,
\begin{equation}
    R^{(r)}_{AB}(\mbox{\boldmath $\omega_{\rm V}$};
                 \mbox{\boldmath $\omega_{\rm A}$})
 = \delta_{AB} -\omega_V^{a}\left(T_V^{(r)}\right)^a_{AB}
               -\omega_A^{a}\left(T_A^{(r)}\right)^a_{AB} +\rmO(\omega^2),
\end{equation}
defines the anti-hermitian generators of infinitesimal
$\SUtwo\times\SUtwo$ rotations in the representation $r$.

\section{Equivalence between the renormalized theories}

We discuss under which conditions 
the identity~(\ref{identity_bare}) holds
in terms of renormalized correlation functions.
Without loss one may restrict attention
to correlation functions of composite fields which 
keep a physical distance from each other.

\subsection{Renormalized tmQCD}

So far we have dealt with the theory at fixed lattice
spacing $a$, and it is not obvious that the theory
has a continuum limit. We assume that infrared divergences
of the correlation functions are properly taken care of 
e.g.~by working in a finite volume with
a suitable choice of boundary conditions. Note that this 
also ensures analyticity of the correlation functions 
in the mass parameters. However, we expect that our conclusions
will be valid more generally.

In perturbation theory, it has been  
shown that lattice QCD with Ginsparg-Wilson quarks is 
renormalizable~\cite{Reisz_Rothe}, and we shall assume that
this remains true beyond perturbation theory. 
While this result has been obtained for massless QCD, 
we do not expect any additional complication here,
as both twisted and standard mass terms can be viewed 
as super-renormalizable interaction terms 
which do not modify the power counting.

The entire physical information of QCD is contained in
the correlation functions of gauge invariant composite
fields. Having introduced the lattice regularized
theory, we may completely avoid gauge fixing,
and we may also avoid the renormalization of the
fundamental fields, which only play the r\^ole of integration
variables. We are thus left with the renormalization of
the bare parameters of the action, and the renormalization
of the composite fields which enter the correlation functions.
The symmetries of the lattice regularized theory 
are the same as in the continuum except for the continuous
space-time symmetry being replaced by the symmetry group
of the hyper-cubic lattice. It then follows 
that renormalized parameters take the form, 
\begin{equation}
  \gr^2 = Z_g g_0^2,\qquad
  \mr   = Z_m  m,\qquad
  \mur  = Z_\mu \muq,
\end{equation}
where the renormalization constants are functions, 
\begin{equation}
  Z=Z(g_0^2,a\mu, \muq/\mu,m/\mu),
  \label{ren_const}
\end{equation}
and $\mu$ denotes the renormalization scale.

Even though chiral symmetry is also broken by the
mass terms it is customary to renormalize operators
such that the chiral multiplet structure of the massless
theory carries over to the renormalized theory. 
Renormalized operators then are of the form
\begin{equation}
  (\phi_{\rm R})^{(r)}_{kA}=Z_k\left( \phi^{(r)}_{kA} + 
   c^{(r,r')}_{kk';AA'}a^{d_{k'}-d_k^{}} \phi^{(r')}_{k'A'}
                              \right),
  \label{op_renorm}
\end{equation}
where $d_k^{}$ and $d_{k'}$ are the mass dimensions of the
fields in the multiplets $k$ and $k'$. 
The structure of this equation follows from the well-known
result of renormalization theory which states that composite fields mix
under renormalization with all fields of equal or lower 
dimension, which transform identically under all the symmetries
of the regularized theory. We assume here that $d_{k'}<d_k^{}$,
i.e.~either there is no mixing with fields of equal mass dimensions,
or this has already been taken into account by choosing a basis
where the renormalization matrix is diagonal at the renormalization 
scale $\mu$. Hence the $c$-coefficients in eq.~(\ref{op_renorm})
only multiply operators with lower dimensions, and this implies that 
they cannot depend on the renormalization scale $\mu$~\cite{Testa}.
While the multiplicative renormalization constants $Z_k$ are 
of the form~(\ref{ren_const}), the $c$-coefficients may thus be considered
functions of the bare parameters, $g_0^2,\,am$ and $a\muq$. 
Note that the multiplet structure of the bare theory is respected
by the assignment of a common renormalization constant to all 
members of a multiplet.

In the following we shall assume that the renormalization of
the theory and the composite fields works out along these
lines. While this is guaranteed in perturbation theory,
the non-perturbative renormalization of power divergent operators
may require an additional effort. In particular, it may be
necessary to first implement Symanzik's improvement programme
to sufficiently high order in the lattice spacing $a$ before
the power divergences can be subtracted in an unambiguous way.

\subsection{A special choice of renormalization scheme}

We would like to identify renormalization schemes 
where the equation~(\ref{identity_bare}) carries
over to the renormalized theory,  with 
renormalized fields of the form~(\ref{op_renorm}).
First of all, we notice that the exact PCAC and PCVC relations of the
bare theory imply that $\muq$ and $m$ can 
be renormalized by the same renormalization
constant. We may thus choose $Z_m=Z_\mu$ implying a
multiplicative renormalization of the bare radial mass $M$
by the same constant. Furthermore, $\alpha$ is not
renormalized as it is determined by the ratio of the mass parameters.
As the effect of the chiral rotation is a change
of $\alpha$, we would like to choose all multiplicative 
renormalization constants independently of this angle, i.e.
\begin{equation}
  Z=Z(g_0^2,a\mu,M/\mu).
 \label{Z_radial}
\end{equation}
A simple example is a mass-independent renormalization scheme
which is obtained by renormalizing the theory in the chiral 
limit~\cite{Weinberg}. As we have assumed an infrared regularization
to be in place, such renormalization schemes do exist and 
it is then obvious that eq.~(\ref{identity_bare}) holds for
the renormalized correlation functions of multiplicatively 
renormalizable operators, viz.
\begin{eqnarray}
  \lefteqn{\left\langle 
    (\phi_{\rm R})^{(r_1)}_{k_1A_1}(x_1)\cdots
    (\phi_{\rm R})^{(r_n)}_{k_nA_n}(x_n)
  \right\rangle_{(M_{\rm R},\alpha)}   = }\nonumber\\
  &&\left\{\prod_{i=1}^n R_{A_iB_i}^{(r_i)}(-\alpha')\right\}
  \left\langle
    (\phi_{\rm R})^{(r_1)}_{k_1B_1}(x_1)\cdots
    (\phi_{\rm R})^{(r_n)}_{k_nB_n}(x_n)
  \right\rangle_{(M_{\rm R},\alpha-\alpha')}.  
 \label{meq}
\end{eqnarray}
The expectation values in the renormalized theory are defined
as in the bare theory, except that the bare parameters
are expressed in terms of the renormalized parameters
$M_{\rm R}$ and $\gr$. Note that eq.~(\ref{meq}) is again
an exact identity, if this is true for~(\ref{identity_bare}) 
and provided the renormalization constants 
are chosen exactly as specified above (i.e.~not only up to 
cutoff effects).

The case of power divergent operators is slightly more
complicated. In general, if eq.~(\ref{meq}) is to be satisfied
the c-coefficients cannot be independent of $\alpha$. 
For example, the renormalized scalar and pseudo-scalar
densities are of the form
\begin{eqnarray}
  i(P_{\rm R})^a &=& \Zp\left(iP^a + \delta^{a3} a^{-3}c_{\rm P}\right),
  \label{axial_ren}\\ 
  \frac12 (S_{\rm R})^0 &=&
       Z_{\rm S}\left( \frac12 S^0 + a^{-3} c_{\rm S}\right),
   \label{scalar_ren}
\end{eqnarray}
with  $Z_{\rm S}=Z_{\rm P}$, and it is well-known 
that $c_{\rm P}$ vanishes at $\muq=0$.

Rather than being independent of $\alpha$, the additive
counterterms satisfy a covariance condition. 
More precisely, assuming that all multiplicative renormalization constants
are of the form~(\ref{Z_radial}),
the requirement that eqs.~(\ref{identity_bare}) and (\ref{meq}) 
hold simultaneously implies,
\begin{equation}
   R_{AB}^{(r)}(-\alpha') c_{kk';BC}^{(r,r')}(g_0^2,aM,\alpha-\alpha') =
   c_{kk';AB}^{(r,r')}(g_0^2,aM,\alpha) R_{BC}^{(r')}(-\alpha'), 
  \label{covariance}
\end{equation}
where only the index $B$ is summed, all others being fixed.
In the example of the pseudo-scalar and scalar densities one finds 
the equations
\begin{equation}
   \begin{pmatrix}
     c_{\rm P}(\alpha) \\ 
     c_{\rm S}(\alpha) 
   \end{pmatrix}
   =
   \begin{pmatrix}
     \cos\alpha' & \sin\alpha'\\
     -\sin\alpha'& \cos\alpha'
   \end{pmatrix}
   \begin{pmatrix}
     c_{\rm P}(\alpha-\alpha') \\ 
     c_{\rm S}(\alpha-\alpha') 
   \end{pmatrix},
\end{equation}
where we have only indicated the dependence upon the angle. It is
obvious that  $c_{\rm P}^2+  c_{\rm S}^2$ is independent
of $\alpha$ and due to the vanishing of  $c_{\rm P}(0)$,
the solution can be parameterised as follows,
\begin{equation}
  c_{\rm P}(\alpha)=c_{\rm S}(0)\sin(\alpha),\qquad
  c_{\rm S}(\alpha)=c_{\rm S}(0)\cos(\alpha).
\label{add_example}
\end{equation}
In other words, if the renormalization problem for the
scalar density can be  solved in standard QCD, the solution
at all other values of $\alpha$ is given by these equations.

\subsection{Differential equation in $\alpha$}

The requirements for the renormalization scheme are best 
characterised by a differential equation in $\alpha$.
We start by considering the third flavour component of the
axial Ward identity~(\ref{AWI}), chose the region 
$R$ to be the space-time manifold
itself and obtain the {\em exact} identity of the bare theory,
\begin{eqnarray}
  \label{alpha_der_bare} 
  \lefteqn{\left.{\partial \over \partial \alpha }\right|_{M,g_0} 
   \left\langle   
   \prod_{i=1}^n \phi^{(r_i)}_{k_i A_i}(x_i)  
   \right\rangle_{(M,\alpha)}=}\nonumber\\ 
  &&\sum_{i=1}^n T^{(r_i)}_{A_iB}\left\langle 
    \phi_{k_1A_1}^{(r_1)}(x_1)\cdots\phi_{k_iB}^{(r_i)}(x_i)\cdots
    \phi_{k_nA_n}^{(r_n)}(x_n)\right\rangle_{(M,\alpha)},
\end{eqnarray}
where $T^{(r)}\equiv\bigl(T_A^{(r)}\bigr)^3$. 
Note in particular that the normalising factor of the path integral is
independent of $\alpha$ and hence does not generate a disconnected
contribution.

Next we recall that analyticity of the renormalized 
correlation functions in the mass parameters 
$\mr$ and $\mur$ is guaranteed by the
infrared cutoff. Therefore, also their derivative 
with respect to $\alpha$ must exist, as the relation
\begin{equation}
   \left.{\partial \over {\partial \alpha}}\right|_{M_{\rm R},g_{\rm R}}
  = \mr \left.{{\partial}\over{\partial\mur}}\right|_{\mr,g_{\rm R}}
 -\mur \left.{{\partial}\over{\partial\mr}}\right|_{\mur,g_{\rm R}},
\end{equation}
follows directly from the definition of the (renormalized) polar mass
coordinates~[cf. eq.~(\ref{polar}]. 
Differentiating a renormalized $n$-point function
at fixed renormalized parameters, 
applying the chain rule and using eq.~(\ref{alpha_der_bare}),
one obtains the  differential equation,  
\begin{eqnarray}
  \label{A3_rep_inv}
  \lefteqn{\left(\nabla_\alpha -\sum_{i=1}^n l_{k_i} \right)\left\langle 
    (\phi_{\rm R})^{(r_1)}_{k_1 A_1}(x_1)\cdots 
    (\phi_{\rm R})^{(r_n)}_{k_n A_n}(x_n)
  \right\rangle_{(M_{\rm R}, \alpha)}=}  \nonumber\\
  &&\!\!\! \sum_{i=1}^n T_{A_i B}^{(r_i)} 
  \left\langle  (\phi_{\rm R})^{(r_1)}_{k_1 A_1}(x_1)\cdots
  (\phi_{\rm R})^{(r_i)}_{k_i B}(x_i) \cdots 
   (\phi_{\rm R})^{(r_n)}_{k_n A_n}(x_n)
  \right\rangle_{(M_{\rm R}, \alpha)}\!\!.
\end{eqnarray}
Here the differential operator $\nabla_\alpha$ is defined through
\begin{equation}
 \nabla_\alpha \defeq  
   \left.{\partial \over \partial \alpha}\right|_{M_{\rm R},g_{\rm R}} +
    l_{\rm M} M_{\rm R} 
    \left.{\partial \over \partial M_{\rm R}}\right|_{\alpha,g_{\rm R}} +
    l_{\rm g} g^2_{\rm R} 
    \left.{\partial \over \partial g^2_{\rm R}}\right|_{\alpha,M_{\rm R}}, 
\end{equation}
and the coefficients are given by 
\begin{equation}
  \label{l_coeff}
  l_{\rm X} = \left.{\partial \log Z_{\rm X}\over \partial \alpha}
  \right|_{M,g_0},\qquad {\rm X}=M,g,k_1,\ldots,k_n.  
\end{equation}
With a suitable choice of the irrelevant parts of the
renormalization counterterms eq.~(\ref{A3_rep_inv}) holds exactly
at finite lattice spacing. The very existence of the differentiated 
correlation function then leads to a constraint for the additive
counterterms, viz.
\begin{equation} 
   \left.{\partial \over \partial \alpha }\right|_{M,g_0} 
  c_{kk';AC}^{(r,r')}   = 
   \left\{T_{AB}^{(r)} c_{kk';BC}^{(r,r')} -c_{kk';AB}^{(r,r')} T_{BC}^{(r')} 
  \right\}.
  \label{additive}
\end{equation}
As one might expect, integrating  
eq.~(\ref{additive}) from $\alpha$ to $\alpha-\alpha'$ reproduces
the covariance relation (\ref{covariance}). Note however,
that eq.~(\ref{additive}) holds without assuming~eq.~(\ref{Z_radial}).

To relate the renormalized correlation functions 
defined at two different values of the 
angle $\alpha$ one just has to integrate
the differential equation~(\ref{A3_rep_inv}). It is possible to 
formally write down the general solution. However, we do not
find this particularly illuminating and therefore just mention that
the general relation is much more complicated than eq.~(\ref{meq}).
The question then arises under which conditions
the relation does take the simple form~(\ref{meq}).
For this to work out, the l.h.s.~of eq.~(\ref{A3_rep_inv}) should  
reduce to the partial derivative with respect to $\alpha$, 
i.e.~one requires
\begin{eqnarray}
 \lefteqn{\left[l_{\rm M} M_{\rm R} 
    \left.{\partial \over \partial M_{\rm R}}\right|_{\alpha,g_{\rm R}} +
    l_{\rm g} g^2_{\rm R} 
    \left.{\partial \over \partial g^2_{\rm R}}\right|_{\alpha,M_{\rm R}}
   -\sum_{i=1}^n l_{k_i}\right]} \nonumber\\
  &&\times\left\langle 
    (\phi_{\rm R})^{(r_1)}_{k_1 A_1}(x_1)\cdots 
    (\phi_{\rm R})^{(r_n)}_{k_n A_n}(x_n)
  \right\rangle_{(M_{\rm R}, \alpha)}=0.
\end{eqnarray}
Note that such an equation must hold for all
renormalized correlation functions. Therefore, 
unless $M_{\rm R}$ and $\gr$ are related in a special way, 
the $l_{\rm X}$ must all vanish, i.e.~the 
multiplicative renormalization constants all take the
form~(\ref{Z_radial}).

We finally remark that eq.~(\ref{A3_rep_inv}) 
expresses the re-parameterisation invariance of the theory:
an infinitesimal change of quark variables of the 
form~(\ref{varchange}) is compensated
by a change of $\alpha$, $M_{\rm R}$ and $g_{\rm R}$.
Our derivation makes use of the axial Ward identity 
as the change of variables~(\ref{varchange}) 
corresponds to a special chiral symmetry transformation.
However, more general changes of variables 
can be considered using rather general 
results of renormalization theory~\cite{RepInV_pap}. 

\subsection{The r\^ole of the Ward identities}

Eq.~(\ref{meq}) is the relation between renormalized tmQCD 
and QCD in its simplest form. Its infinitesimal version
is eq.~(\ref{A3_rep_inv}) with vanishing 
coefficients $l_{\rm X}$~(\ref{l_coeff}). 
We have argued that it is possible to renormalize the
Ginsparg-Wilson regulated theory such that eq.~(\ref{meq})
holds, and we have worked out the conditions on the
renormalization constants in this regularization.
Based on universality one thus expects that tmQCD in other,
not necessarily chirally invariant regularizations
can again be renormalized such that eq.~(\ref{meq}) is satisfied
up to cutoff effects. At least in perturbation theory 
this can be proved rigorously.

In view of the formulation of tmQCD with Wilson quarks, 
we would like to emphasise the r\^ole of the renormalized
tmQCD Ward identities, which are of the same form as 
eqs.~(\ref{AWI},\ref{VWI}).
As is well-known, the Ward identities ensure
that the renormalized composite fields form chiral multiplets,
and fix the absolute normalization of the symmetry currents.
Furthermore, the mass parameters are renormalized inversely
to the pseudo-scalar and scalar densities, and this 
implies that one may set
\begin{equation}
   \mr=M_{\rm R}\cos\alpha,\qquad \mur=M_{\rm R} \sin\alpha,
\end{equation}
with the angle $\alpha$ which remains unrenormalized.

We are now going to demonstrate that the 
renormalized tmQCD Ward identities also imply that
certain linear combinations of the correlation functions 
satisfy the Ward identities of standard QCD. 
In the following we assume the continuum limit has
been taken and use a shorthand notation for the
renormalized  correlation functions,
\begin{equation}
    G\equiv 
   \left\langle (\phi_{\rm R})_{kA}^{(r)}(y) {\cal O}_{\rm ext}
   \right\rangle_{(M_{\rm R},\alpha)},
\end{equation}
where ${\cal O}_{\rm ext}$ is some product of renormalized
fields which are localized in the exterior of the finite
physical space-time region $R$. Furthermore we assume $y\in R$
and denote by $G^a_{\rm A,V,P}$ and $G_{\rm S}$
the same correlation functions with an insertion of 
$\partial_\mu (A_{\rm R})_\mu^a$, $\partial_\mu (V_{\rm R})_\mu^a$,
$(P_{\rm R})^a$ and $(S_{\rm R})^0$ at the space-time point $x$,
integrated over the region $R$ with respect to $x$.
The renormalized Ward identities may then be written in the form
\begin{eqnarray}
  G_{\rm A}^a-2M_{\rm R}\left\{\cos(\alpha) G_{\rm P}^a
      +i\sin(\alpha) \delta^{3a}G_{\rm S}\right\} 
   &=& -iT_{\rm A}^a G,\\[2ex]
  G_{\rm V}^a+2M_{\rm R}\sin(\alpha)\varepsilon^{3ab} G^b_{\rm P} 
   &=& -iT_{\rm V}^a G,
\end{eqnarray}
where the abbreviation
\begin{equation}
  T_{\rm X}^a G\equiv 
   \left(T_X^{(r)}\right)^a_{AB}
  \left\langle (\phi_{\rm R})_{kB}^{(r)}(y) {\cal O}_{\rm ext}
   \right\rangle_{(M_{\rm R},\alpha)}
\end{equation}
has been used for ${\rm X}={\rm A,V}$.

We now consider the two linear combinations
\begin{eqnarray}
  {\rm c}G_{\rm A}^1 +{\rm s}G_{\rm V}^2 -2M_{\rm R}\left\{{\rm c}\cos\alpha+
  {\rm s}\sin\alpha\right\}G_{\rm P}^1 &=& 
  -i\left({\rm c}T_{\rm A}^1+{\rm s}T_{\rm V}^2\right)G,\\[1ex]
  {\rm c}G_{\rm V}^2 -{\rm s}G_{\rm A}^1 -2M_{\rm R}\left\{{\rm c}\sin\alpha-
  {\rm s}\cos\alpha\right\}G_{\rm P}^1 &=& 
  -i\left({\rm c}T_{\rm V}^2-{\rm s}T_{\rm A}^1\right)G.
\end{eqnarray}
Multiplying both sides of the equation by the 
matrices $R^{(r)}(\alpha)$, setting 
\begin{equation}
  R_\alpha G\equiv R^{(r)}_{AB}(\alpha) 
  \left\langle (\phi_{\rm R})_{kB}^{(r)}(y) {\cal O}_{\rm ext}
   \right\rangle_{(M_{\rm R},\alpha)}
\end{equation}
and also defining
\begin{equation}
  G_{\rm A'}^1\equiv{\rm c}G_{\rm A}^1 +{\rm s}G_{\rm V}^2,\qquad
  G_{\rm V'}^2\equiv{\rm c}G_{\rm V}^2 -{\rm s}G_{\rm A}^1,
\end{equation}
we then find 
\begin{eqnarray}
  R_\alpha G_{\rm A'}^1-2M_{\rm R}\left\{{\rm c}\cos\alpha+
  {\rm s}\sin\alpha\right\}R_\alpha G_{\rm P}^1 \! &=& \! 
  -iR_\alpha \left({\rm c}T_{\rm A}^1+{\rm s}T_{\rm V}^2\right)G,\\[1ex]
  R_\alpha G_{\rm V'}^2 -2M_{\rm R}\left\{{\rm c}\sin\alpha-
  {\rm s}\cos\alpha\right\}RG_{\rm P}^1 \! &=&\! 
  -iR_\alpha \left({\rm c}T_{\rm V}^2-{\rm s}T_{\rm A}^1\right)G.
\end{eqnarray}
We notice that in order to preserve the canonical normalization of
the primed currents, one needs ${\rm c}^2+{\rm s}^2=1$, i.e.~we may set
\begin{equation}
   {\rm c}=\cos(\alpha'),\qquad {\rm s}=\sin(\alpha').
\end{equation}
It is then clear that the standard Ward identities can be obtained
by choosing $\alpha'=\alpha$, and provided the equations
\begin{eqnarray}
  R^{(r)}_{AB}(\alpha)
  \left({\rm c}\bigl(T^{(r)}_{\rm A}\bigr)^1_{BC}
       +{\rm s}\bigl(T^{(r)}_{\rm V}\bigr)^2_{BC}\right)
      R^{(r)}_{CD}(-\alpha)
  &=& \bigl(T^{(r)}_{\rm A}\bigr)^1_{AD}, \label{rot1}\\
  R^{(r)}_{AB}(\alpha)
  \left({\rm c}\bigl(T^{(r)}_{\rm V}\bigr)^2_{BC}
       -{\rm s}\bigl(T^{(r)}_{\rm A}\bigr)^1_{BC}\right)
        R^{(r)}_{CD}(-\alpha)
  &=& \bigl(T^{(r)}_{\rm V}\bigr)^2_{AD},\label{rot2}
\end{eqnarray}
hold. By differentiating
with respect to $\alpha$ and using the Lie algebra of the 
$\SUtwo\times\SUtwo$ generators,
\begin{equation}
   \bigl[T_{\rm A}^a, T_{\rm A}^b\bigr] = \varepsilon^{abc} T_{\rm V}^c,
    \quad
   \bigl[T_{\rm A}^a, T_{\rm V}^b\bigr] = \varepsilon^{abc} T_{\rm A}^c,
    \quad
   \bigl[T_{\rm V}^a, T_{\rm V}^b\bigr] = \varepsilon^{abc} T_{\rm V}^c,
\end{equation}
one arrives at the conclusion that eqs.~({\ref{rot1},\ref{rot2}) are
indeed satisfied. Note that the same procedure applies to the 
remaining flavour components of the Ward identities.
Hence, the validity of the tmQCD Ward identities implies that
\begin{itemize}
 \item there exist particular linear combinations  
  of correlation functions which satisfy the standard QCD Ward identities 
 for two degenerate quark flavours with mass $M_{\rm R}$,
 \item the linear combinations only depend on the angle $\alpha$, which
 is determined  by the ratio of the quark 
 mass parameters which appear in the Ward identities.
\end{itemize}
In particular it is clear that the angle $\alpha$ has no physical
significance. We may start with any value
of $\alpha$ and still obtain the standard chiral flavour Ward identities.
As a given theory is identified by its symmetries this 
implies the equivalence of theories defined at different values of 
$\alpha$, provided the remaining symmetries are 
either $\alpha$-independent or transform covariantly. 
This is certainly true on the level 
of the renormalized composite fields: for any transformation
of the composite fields at $\alpha=0$ one may identify the corresponding
transformation in the renormalized twisted theory. 

\subsection{Concluding remarks}

In practical applications one would like to work with tmQCD
at a given fixed value of $\alpha$, and just invoke
eq.~(\ref{meq}) in order to {\em interpret} the results
in terms of standard QCD. According to the above
discussion one may start by imposing the tmQCD Ward 
identities in the renormalized theory. Besides defining the 
value of $\alpha$ this procedure restores the chiral multiplet
structure of the bare theory. One then still needs to 
impose a renormalization condition per chiral multiplet.
If this is done either in the chiral limit or 
independently of $\alpha$, eq.~(\ref{meq}) provides the
relation to the theory defined at any other angle $\alpha$, 
including $\alpha=0$. The simplification in the Ginsparg-Wilson
regularization consists in the validity of 
bare continuum-like tmQCD Ward identities,
and in the related fact that the bare Ward identity
masses coincide with the bare mass parameters of the action.
Finally we stress that the classical continuum theory
allows to infer the relation~(\ref{meq}) between renormalized theories 
and may hence be used as a guide.

\section{Twisted mass QCD with Wilson quarks}

In this section we discuss in some detail the
regularization with Wilson fermions, including some
practical aspects of applications.

\subsection{Symmetries of the bare theory}

With Wilson quarks the tmQCD Dirac operator
is as given in eq.~(\ref{D_tmQCD}) with the usual massless
Wilson-Dirac operator\footnote{For unexplained
notation and conventions we refer to ref.~\cite{paperI}}
\begin{equation}
  D_W=\frac12\sum_{\mu=0}^3\Bigl\{\dirac{\mu}(\nab{\mu}+\nabstar{\mu})
             -a\nabstar{\mu}\nab{\mu}\Bigr\}.
\label{WilsonDirac}
\end{equation}
For simplicity we defer the discussion of O($a$) improved tmQCD
to a separate publication~\cite{II}. Here we note that the 
Wilson term is not left invariant by the axial rotation~(\ref{axial}), 
and the lattice regulated theories at $\muq=0$ and $\muq\ne 0$ 
are thus different. This is of course
welcome as otherwise the zero mode problem would be
present in both cases~(cf.~sect~1). One may think of more
general lattice Dirac operators, also including a chirally twisted
Wilson term. However, a moment of thought reveals that this
is not really more general, as an axial rotation~(\ref{axial}) may then
be used in the lattice theory to eliminate the extra
Wilson term. Modulo a more general coefficient of the standard
Wilson term and with re-defined bare mass parameters, one then obtains
again the action corresponding to eq.~(\ref{D_tmQCD}).

As compared to the theory with $\Nf=2$ standard 
Wilson quarks ($\muq=0$) we find that the 
exact U(2) symmetry is reduced to a U(1)
symmetry leading to fermion number conservation, and 
a vectorial U(1) symmetry with generator $\tau^3/2$. 
Concerning the space-time symmetries,
the lattice action is invariant under axis permutations, whereas
reflections such as parity are a symmetry
only if combined with either a flavour exchange
\begin{equation}
  F:\qquad \psi    \rightarrow \tau^1 \psi,
    \qquad \psibar \rightarrow \psibar \tau^1,
\end{equation}
or a sign change of the twisted mass term $\muq \rightarrow -\muq$.
We will refer to the thus modified parity symmetries as $P_F$ and $\tilde P$
respectively. The list of exact symmetries is completed by charge
conjugation, and we note that twisted mass lattice QCD with Wilson quarks
has a positive self-adjoint transfer matrix~\cite{II}. 

\subsection{Renormalized parameters}

As in sect.~3 we assume that infrared divergences are
regulated e.g. by working in a finite space-time volume
with suitable boundary conditions. This implies analyticity
of the theory in the mass parameters and it is then 
rather obvious that twisted mass lattice QCD 
is renormalizable by  power counting~\cite{Reisz}. The counterterm
structure follows from the symmetries of the regularization.
Based on these symmetries one concludes that tmQCD is 
finite after the usual renormalization of 
the coupling and the standard mass parameter, 
\begin{equation}
  \gr^2 =Z_g g_0^2,\qquad \mr=Z_m \mq, \qquad \mq=m_0-\mc, 
 \label{param_renorm}
\end{equation}
and, in addition a {\em multiplicative} 
renormalization of the twisted mass parameter,
\begin{equation}
  \mur =Z_\mu \muq.
\end{equation}
In particular we note that the modified parity symmetry, $P_F$,
is sufficient to exclude a counterterm 
$\propto \tr\{F_{\mu\nu} \tilde{F}_{\mu\nu}\}$.

\subsection{Ward identities and renormalization of composite fields}

Following sect.~3 we require that the
renormalized theory satisfies the axial and vector Ward 
identities, i.e.~the renormalized analogues of 
eqs.~(\ref{AWI},\ref{VWI}). Concerning the vector
Ward identity the situation is the same as with Ginsparg-Wilson
fermions, due to the exact flavour symmetry of massless
Wilson quarks. Therefore eq.~(\ref{VWI}) holds exactly,
with the point-split vector current
\begin{eqnarray}
   \widetilde{V}^a_\mu(x) &=&  
 \frac12 \Bigl\{\psibar(x)(\gamma_\mu-1)\frac{\tau^a}{2}
 U(x,\mu)\psi(x+a\hat\mu) \nonumber\\
 && \hphantom{\frac12 } +\psibar(x+a\hat\mu)(\gamma_\mu+1)\frac{\tau^a}{2} 
 U(x,\mu)^{-1}\psi(x)\Bigr\},
\end{eqnarray}
and the local pseudo-scalar density. It then follows that the
vector current is protected against renormalization, i.e.~$\Zv=1$.
More generally, the multiplicative renormalization constants of
composite fields which belong to the same isospin multiplet 
must be identical in order to respect the vector Ward identities. 
An example is the renormalized pseudo-scalar
density which has the structure
\begin{equation}
  i(P_{\rm R})^a =
     \Zp\left\{iP^a+\delta^{a3} a^{-3} c_{\rm P}\right\}
\end{equation}
and $c_{\rm P}$ vanishes exactly at $\muq=0$. The vector Ward
identity here implies that $\Zp$ is the same for all flavour
components, and, moreover, 
\begin{equation}
  \Zp=Z_\mu^{-1}.
  \label{Zmu}
\end{equation}
In contrast, the axial Ward identity does not hold in the
bare theory. Axial Ward identities therefore provide 
normalization conditions which determine finite renormalization
constants such as $\Za$, or finite 
ratios of scale dependent renormalization constants,
such as $Z_{\rm S}/Z_{\rm P}$~\cite{Bochicchio_et_al}.
Moreover, these finite renormalization constants only depend
on the bare coupling $g_0$ and may therefore be determined
in the chiral limit using standard procedures~\cite{paperIV}.
Note that the finite renormalization constants 
restore chiral symmetry of the bare theory up to cutoff effects. 
Once this is achieved the renormalization
of multiplicatively renormalizable fields is similar to the
Ginsparg-Wilson case, i.e.~the renormalization constant for a
given multiplet is determined by imposing a renormalization 
condition on one of its members. Of particular practical 
interest are mass-independent renormalization schemes,
which are obtained by imposing a renormalization condition
at $\muq=\mq=0$~\cite{Weinberg}. Based on universality 
we then expect that the relations~(\ref{meq}) between renormalized 
correlation functions hold up to cutoff effects. According to
sect.~3 the same can be achieved by imposing $\alpha$-independent
renormalization conditions, where, in the case of Wilson fermions
the angle $\alpha$ must be defined through the Ward identity 
masses (cf. subsect.~4.4).

In principle the Ward identities also determine additive
renormalization constants which arise due to 
the explicit breaking of chiral symmetry. An example
is the renormalization of the iso-singlet scalar density, which
has the same structure as in eq.~(\ref{scalar_ren}), however, 
with a coefficient $c_{\rm S}$ which does not vanish in the
chiral limit. Therefore, the renormalization of 
the third axial Ward identity~(\ref{AWI}) 
requires the explicit subtraction of power divergences. 
While power divergent renormalization problems
do not present any particular difficulty in perturbation theory,
it is less clear how to proceed in a non-perturbative approach.
A general discussion of this topic is beyond the scope of
this work. Here we just note that
the renormalization of the third axial Ward identity 
may in fact be avoided if one assumes that the physical 
isospin symmetry is restored in the renormalized
theory. In the following we will make this (plausible) assumption and
not discuss the third axial Ward identity any further.

\subsection{Definition of the angle $\alpha$}

According to section~3.4 the angle $\alpha$ is uniquely defined
through the renormalized Ward identity masses. 
Assuming that $\Zp$ has been fixed, the renormalized twisted
mass parameter is determined due to eq.~(\ref{Zmu}).
The renormalized axial current and the first two components
of the pseudo-scalar density may then be used to 
define $\mr$ through the renormalized PCAC relation
\begin{equation}
  \partial_\mu (A_{\rm R})_\mu^a=2\mr(P_{\rm R})^a, \qquad a=1,2.
\end{equation}
In practice one first defines a bare PCAC mass $m$ 
from the ratio of correlation functions involving the bare
axial current and pseudo-scalar density. 
The renormalized PCAC mass is then given by
\begin{equation}
  \mr=\Za Z_{\rm P}^{-1} m = Z_{\rm m}\mq.
\end{equation}
Using eq.~(\ref{Zmu}), the angle $\alpha$ is then determined as 
\begin{equation}
   \tan\alpha= {\mur\over\mr} = {\muq\over{\Za m}}
  ={\muq\over{Z_{\rm m}\Zp\mq}}.
\end{equation}
In general one thus needs the bare PCAC mass $m$ and the
axial current normalization constant $\Za$
to obtain $\alpha$. Note that the latter is not needed in the 
special case $m=0$, i.e.~if $\alpha=\pi/2$.

\subsection{Avoiding lattice renormalization problems}  

Once tmQCD has been renormalized in the way described above, 
eq.~(\ref{meq}) can be applied to establish a ``dictionary'' 
between the renormalized correlation functions in QCD and tmQCD.
For example, the 2-point function of the axial current and the
pseudo-scalar density in standard QCD translates as follows 
\begin{eqnarray}
  \left\langle (A_{\rm R})_0^1(x)
   (P_{\rm R})^1(y)\right\rangle_{(M_{\rm R},0)}
  &=&\cos(\alpha)\left\langle (A_{\rm R})_0^1(x)(P_{\rm R})^1(y)
                \right\rangle_{(M_{\rm R},\alpha)}\nonumber\\
  &&\hphantom{} 
  +\sin(\alpha)\bigl\langle \widetilde{V}_0^2(x)(P_{\rm R})^1(y)
                \bigr\rangle_{(M_{\rm R},\alpha)}.
 \label{2pt}
\end{eqnarray}
More generally, relations between the renormalized composite
fields can be inferred from the corresponding relations
in the classical theory~(cf.~sect.~2). In particular,
the above example follows from 
eqs.~(\ref{axial_current_rot}--\ref{axial_density_rot}).

As tmQCD and standard QCD with Wilson quarks are
not related by a lattice symmetry, the
counterterm structure for composite fields with the
same physical interpretation depends upon $\alpha$.
Given a physical amplitude it may hence be possible
that  a particular choice of $\alpha$ leads
to simplifications.
An obvious case is the computation of $F_\pi$ from the
2-point function~(\ref{2pt}). While the standard approach
(i.e.~the direct computation of the l.h.s.) requires 
to first determine the renormalized axial current,
the r.h.s.~of this equation at $\alpha=\pi/2$ only
contains the vector current which is protected against
renormalization.

Even more interesting is the application 
of tmQCD to  matrix elements of the iso-singlet scalar density. 
At $\alpha=\pi/2$, the physical scalar density is 
represented by the third component of the pseudo-scalar 
density, see eq.~(\ref{scalar_density_rot}). 
While the scalar density is cubically 
divergent even in the chiral limit,
the pseudo-scalar density has a quadratic divergence which
vanishes {\em exactly} at $\muq=0$. 
The situation is therefore comparable to the case of the
renormalized scalar density in standard QCD with
Ginsparg-Wilson fermions.

\subsection{Inclusion of heavier quarks}

It is straightforward to generalise tmQCD to include any number of
heavier quark flavours. For the latter one 
may use the standard regularization with (improved) Wilson quarks,
as the zero mode problem is practically absent at mass parameters
which correspond to the physical strange quark mass. The renormalization 
procedure can again be carried out such that the chiral flavour  
Ward identities are respected, which now 
also involve mixed operators of light and heavy quarks. 
Hence we expect that the ``dictionary'' 
between tmQCD and standard QCD can again
be established by naive continuum considerations.
As an example we consider the tmQCD continuum 
action~(\ref{tmQCDcont}) and add the action of the strange quark,
\begin{equation}
  S_{\rm F}[\psi,\psibar,s,\bar{s}] =
  \int \rmd^4x\,\left\{\psibar\left(D\kern-7pt\slash
                +m+i\muq\gamma_5\tau^3\right)\psi 
    +\bar{s}\left(D\kern-7pt\slash+m_s\right)s\right\}.
\label{tmQCDcont_s}
\end{equation}
Using a physical notation,
\begin{equation}
   \psi = \begin{pmatrix} u \\ d  \end{pmatrix}, \qquad 
   \psibar =\begin{pmatrix} \bar{u} & \bar{d}\, \end{pmatrix}, 
\end{equation}
the standard PCAC relation,
\begin{equation}
  \partial_\mu (\bar{d}' \gamma_\mu\gamma_5 s) 
  = (m'+m_s)\, \bar{d}'\gamma_5 s,
\end{equation}
is obtained with the rotated axial current and pseudo-scalar density,
\begin{eqnarray}
  \bar{d}' \gamma_\mu\gamma_5 s &=&  \cos(\frac12\alpha)\, 
        \bar{d}\gamma_\mu\gamma_5 s +i\sin(\frac12\alpha)\,
         \bar{d}\gamma_\mu s,
 \label{interpol1}\\
   \bar{d}'\gamma_5 s           &=& \cos(\frac12\alpha)\, \bar{d}\gamma_5 s 
             -i\sin(\frac12\alpha)\, \bar{d} s,
\label{interpol2}
\end{eqnarray}
and with the angle $\alpha$ and the light quark mass 
$m'$ as given in sect.~2.1.

\subsection{Application to the $\Delta S=2$ effective weak Hamiltonian}

Also in the case of operators involving light and
strange quarks certain renormalization problems of standard Wilson quarks
can be circumvented. An interesting example 
is the $\Delta S=2$ part of the
effective weak Hamiltonian,
\begin{equation}
   O^{\Delta S=2} = \left\{\bar{s}\gamma_\mu(1-\gamma_5)d\right\}^2.
\end{equation}
In phenomenology one is mainly interested in the hadronic matrix element
of this operator between $K_0$ and $\overline{K}_0$ states~\cite{Laurent}. 
As parity does not change in this transition,
only the parity conserving part of the operator contributes. 
Hence one decomposes the operator into parity even and odd parts,
\begin{equation}
  O^{\Delta S=2}=O_{\rm VV+AA}-2O_{\rm VA}.
\end{equation}
In the regularization with Ginsparg-Wilson fermions the operator
$O^{\Delta S=2}$ and thus both $O_{\rm VV+AA}$ and $O_{\rm VA}$ are 
renormalized multiplicatively. With Wilson quarks, 
the remaining symmetries imply that $O_{\rm VV+AA}$ 
mixes with four other parity even operators of 
the same mass dimension, whereas $O_{\rm VA}$ is still renormalized 
multiplicatively~\cite{Bernard_et_al}.
In tmQCD we now observe that
the parity even operator in the standard basis 
is represented by the combination
\begin{equation}
  O'_{\rm VV+AA}=\cos(\alpha) O_{\rm VV+AA} -2i\sin(\alpha)O_{\rm VA}.
\end{equation}
In particular, at $\alpha=\pi/2$, only $O_{\rm VA}$ appears on
the r.h.s., and one concludes that matrix elements of 
the physical operator $O'_{\rm VV+AA}$ can be computed in tmQCD
without solving the complicated renormalization problem 
for the parity even operator\footnote{For an
alternative proposal see ref.~\cite{Martinelli_BK}.}.
In particular, the $K_0-\overline{K}_0$ mixing
amplitude could be extracted from the 3-point function involving
$O_{\rm VA}$ and appropriately rotated interpolating fields
for the kaons~[cf. eqs.~(\ref{interpol1},\ref{interpol2})].

One might be worried that additional counterterms to $O_{\rm VA}$ 
may be  necessary in tmQCD. As there is no such
term at $\muq=0$,  possible counterterms must be accompanied
by at least one power of the twisted mass parameter,
and the flavour structure requires them to be again four-quark operators.
For dimensional reasons such counterterms can only contribute
cutoff effects of the order $a\muq$, and a closer
look shows that the parity even operators
multiplied by $a\muq$ are indeed allowed by the tmQCD symmetries.

\subsection{Technical complications}

The equivalence between tmQCD and standard QCD is a statement
about the renormalized theories in the continuum limit.
When tmQCD is used to define the standard QCD correlation functions,  
some of the physical symmetries are only restored in the continuum limit. 
In particular, this is the case of the flavour symmetry and parity,
which are exact lattice symmetries in standard QCD with Wilson fermions, 
but which are only recovered in the continuum limit if $\alpha\ne 0$.
In practice the problem shows up e.g.~as an ambiguity in the
definition of $\alpha$, which is induced by the usual ambiguity
in the definition of the critical mass $\mc$ by terms of 
O($a$) (or O($a^2$) if the theory is improved)~\cite{paperI}.
To illustrate the consequences  consider
the computation  of the pion mass at fixed cutoff $a$, using
the 2-point function
\begin{equation}
  G(x_0-y_0) = a^3\sum_{\bfx}\langle {P'}^3(x){P'}^3(y)\rangle.
 \label{corr_fct}
\end{equation}
In tmQCD this correlation function is computed by replacing
${P'}^3$ as in eq.~(\ref{axial_density_rot}), 
where the relative multiplicative
renormalization of $S^0$ and $P^3$ is assumed to have been fixed
by the axial Ward identity. Note also that the 
exponential decay of this correlation can be determined
without knowledge of the additive renormalization constants.

It is now obvious that both the ambiguity in $\alpha$
and the O($a$) ambiguity in the relative 
renormalization of the densities imply that the correlation 
function~(\ref{corr_fct}) contains cutoff effects
which are proportional to the propagator of the (physical) scalar
density ${S'}^0$. An analysis of the 2-point function~(\ref{corr_fct})
at fixed $a$ must therefore take into account the states
with the quantum numbers of an iso-singlet scalar state. 
We note in passing that O($a$) improvement does not
alter this situation, as it merely reduces the ambiguities 
to O($a^2$). In particular, at fixed cutoff,
the relevant symmetries remain inexact and the qualitative behaviour 
of the 2-point function~(\ref{corr_fct}) remains unchanged.

In general the analysis of the hadron spectrum in tmQCD
at fixed $a$ must include states which are allowed by
the lattice symmetries but have the ``wrong''
continuum quantum numbers. Although this is 
not a fundamental problem, the analysis is somewhat 
more complicated than in lattice QCD with standard Wilson quarks.
We also note a side effect for the determination 
of hadronic matrix elements. When these are extracted from
tmQCD correlation functions it may not be necessary to 
increase the distances until the desired physical state
is completely isolated. It is sufficient to establish
that contributions from  the excited states with 
the correct continuum quantum numbers are negligible, 
as all other contaminations merely modify the cutoff effects
of the matrix element.

\section{Conclusions}

In this paper we have advocated the use of twisted mass 
QCD with Wilson quarks as an alternative regularization of
QCD with two degenerate light quarks. Using Ginsparg-Wilson
fermions as a tool, we have demonstrated in what sense
tmQCD is equivalent to standard QCD. In particular,
we have clarified under which conditions the relations
between renormalized correlation functions take the simple
form~(\ref{meq}), which is the quantum analogue of
the naive relations derived in the classical continuum theory.

Twisted mass lattice QCD provides a clean field theoretical
solution to the problem of unphysical zero modes.
While our work on tmQCD is motivated by this problem, 
we also observe a few additional benefits. 
In particular we have given examples where 
renormalization problems of lattice operators can be circumvented by
working in the fully twisted theory with $\alpha=\pi/2$. 
For the sake of simplicity we did not discuss O($a$) 
improvement of tmQCD. This topic is deferred to a separate
publication~\cite{II}.

First numerical simulations using (quenched) tmQCD
have already been carried out, and a scaling test in a small volume
has been presented in~\cite{scaling}. It is hoped that
the chiral limit can be approached much more closely in tmQCD
than previously possible with Wilson quarks.
In particular the 
ALPHA collaboration plans to extend the work of~\cite{alpha_hsw}
to much smaller quark masses where (quenched) chiral perturbation theory
should be safely applicable~\cite{Michele_et_al}. 
Furthermore, a project to determine the $K_0-\overline{K}_0$ mixing
amplitude using tmQCD is underway~\cite{BKtmQCD}.
In the future, it will also be interesting to see whether 
numerical simulations of full QCD can benefit 
from using a twisted mass term.

\vskip 1ex

This work is part of the ALPHA collaboration research programme.
We would like to thank M.~Della Morte, M.~L\"uscher, G.C.~Rossi, 
R.~Sommer  and A.~Vladikas for useful discussions.
S.~Sint acknowledges partial support by the European Commission under
grant no.~FMBICT972442. P.A.~Grassi is supported by 
the National Science Foundation (grants no.~PHY-9722083 and PHY-0070787).


\begin{thebibliography}{99}

\bibitem{Wilson} 
K.~Wilson, Phys. Rev. D10 (1974) 2445

\bibitem{Bochicchio_et_al}
M. Bochicchio, L. Maiani, G. Martinelli, G. C. Rossi and M. Testa,
Nucl. Phys. B262 (1985) 331

\bibitem{paperIV}
M. L\"uscher, S. Sint, R. Sommer and H. Wittig, 
Nucl. Phys. B491 (1997) 344 

\bibitem{HMCtrouble}
K. Jansen and R. Sommer, Nucl. Phys. B530 (1998) 185

\bibitem{paperIII}
M. L\"uscher, S. Sint, R. Sommer, P. Weisz and U. Wolff,
Nucl. Phys. B491 (1997) 323

\bibitem{BardeenI}
W. Bardeen, A. Duncan, E. Eichten, G. Hockney and H. Thacker,
Phys. Rev. D57 (1998) 1633

\bibitem{MP}
A. Hoferichter, E. Laermann, V. K. Mitrjushkin, M. M\"uller-Preussker 
and P. Schmidt, Nucl. Phys. B (Proc. Suppl.) 63 (1998) 164

\bibitem{Schierholz}
G. Schierholz, M. G\"ockeler, A. Hoferichter, R. Horsley, D. Pleiter, 
P. Rakow and P. Stephenson,
Nucl. Phys. B (Proc. Suppl.) 73 (1999) 889

\bibitem{SW}
B. Sheikholeslami and R. Wohlert,
Nucl. Phys. B259 (1985) 572

\bibitem{ALPHAexceptional}
M. Guagnelli, J. Heitger, R. Sommer and H. Wittig  (ALPHA Collaboration),
Nucl. Phys. B560 (1999) 465

\bibitem{Aoki}
S. Aoki, Phys. Rev. D30 (1984) 2653

\bibitem{BardeenII}
W. Bardeen, A. Duncan, E. Eichten and H. Thacker,
Phys. Rev. D59 (1999) 014507

\bibitem{lat99} 
R. Frezzotti, P. A. Grassi, S. Sint and P. Weisz,
Nucl. Phys. B (Proc. Suppl.) 83 (2000) 941

\bibitem{GW} 
P.H. Ginsparg and K.G. Wilson, Phys. Rev. D25 (1982) 2649

\bibitem{Hasenfratz}
P. Hasenfratz, Nucl. Phys. B525 (1998) 401

\bibitem{Neuberger}
H. Neuberger, Phys. Lett. B417 (1998) 141 

\bibitem{locality}
P. Hern\'andez, K. Jansen and M. L\"uscher, Nucl. Phys. B552 (1999) 363 

\bibitem{LuscherII}
M. L\"uscher, Nucl. Phys. B549 (1999) 295

\bibitem{LuscherI}
M. L\"uscher, Phys. Lett. B428 (1998) 342

\bibitem{Niedermayer_98}
F. Niedermayer, Nucl. Phys. (Proc. Suppl.)  73 (1999) 105

\bibitem{LuscherIII}
M. L\"uscher, Nucl. Phys. B538 (1999) 515

\bibitem{Kikukawa}
Y. Kikukawa and A. Yamada,
Nucl. Phys. B547 (1999) 413

\bibitem{Reisz_Rothe}
T. Reisz and H. J. Rothe, Nucl. Phys. B575 (2000) 255

\bibitem{Testa}
M. Testa, JHEP  9804 (1998) 002

\bibitem{Weinberg}
S. Weinberg, Phys. Rev. D8 (1973) 3497

\bibitem{RepInV_pap}
R. Frezzotti and P. A. Grassi, in preparation


\bibitem{paperI}
M. L\"uscher, S. Sint, R. Sommer and P. Weisz,
Nucl. Phys. B478 (1996) 365

\bibitem{Reisz}
T. Reisz, Comm. Math. Phys. 116 (1988) 81, 573;
{\it ibid} 117 (1988) 79, 639; Nucl. Phys. B318 (1989) 417

\bibitem{Laurent}
see e.g.~L. Lellouch, {\tt hep-lat/0011088}

\bibitem{Bernard_et_al}
C.~Bernard, T. Draper, G. Hockney and A. Soni,
Nucl. Phys. (Proc. Suppl.) 4 (1988) 483

\bibitem{Martinelli_BK}
D. Becirevic et al., Phys. Lett. B487 (2000) 74


\bibitem{II} 
R. Frezzotti, S. Sint and P. Weisz,
``O($a$) improved twisted mass lattice QCD'', to appear

\bibitem{scaling}
M. Della~Morte, R. Frezzotti, J. Heitger and S. Sint, 
{\tt hep-lat/0010091} and in preparation 

\bibitem{alpha_hsw} J. Heitger, R. Sommer and H. Wittig, 
{\tt hep-lat/0006026}

\bibitem{Michele_et_al}
ALPHA collaboration, work in progress

\bibitem{BKtmQCD} 
ALPHA collaboration \& Rome ``Tor Vergata'', work in progress

\end{thebibliography}
\end{document}